\documentclass[fleqn,10pt]{article}
\usepackage{epsfig}
\newcommand{\be}{\begin{equation}}
\newcommand{\ee}{\end{equation}}

\newcommand{\eq}[1]{eq.~(\ref{#1})}

\newcommand{\chii}{\chi_{{}_{\rm I}}}
\newcommand{\chir}{\chi_{{}_{\rm R}}}
\newcommand{\pbar}{\bar p}

\def\spai{\sigma_{p{-}\rm air}^{\rm inel}}

\begin{document}
\setcounter{secnumdepth}{4}
\renewcommand\thepage{\ }
%
%
\begin{titlepage} 
%
\newcommand\reportnumber{713} 
\newcommand\mydate{December 2000} 
\newlength{\nulogo} 
\settowidth{\nulogo}{\small\sf{University of Wisconsin-Madison:}}
\title{
\vspace{-.8in} 
\hfill\fbox{{\parbox{\nulogo}{\small\sf{Northwestern University: \\
Report No. N.U.H.E.P. \reportnumber\\[1ex]
University of Wisconsin-Madison:\\
Report No. MADPH-00-1203\\[1ex]
          \mydate}}}}
\vspace{0.5in} \\
{
Survival Probability of Large Rapidity Gaps\\ in $\pbar p,\  pp,\ \gamma  
p\ {\rm and \ }\gamma\gamma$ Collisions
}}

\author{
M.~M.~Block
\thanks{Work partially supported by Department of Energy contract
DA-AC02-76-Er02289 Task D.}\vspace{-5pt}   \\
{\small\em Department of Physics and Astronomy,} \vspace{-5pt} \\ 
{\small\em Northwestern University, Evanston, IL 60208}\\
\vspace{-5pt}
\  \\
F.~Halzen
\thanks{Work partially supported by Department of Energy
grant DE-FG02-95ER40896 and the University of Wisconsin Research
Committee with funds granted by the Wisconsin Alumni Research
Foundation.}
\vspace{-5pt} \\
{\small\em Department of Physics,}
\vspace{-5pt} \\
{\small\em University of
Wisconsin, Madison, WI 53706} \\
\vspace{-5pt}\\
%
\vspace{-5pt}\\
%
}    
\vspace{.5in}
\date {}
\maketitle
\begin{abstract}
 Using an eikonal analysis, we simultaneously fit a QCD-inspired  
parameterization
 of all accelerator data on forward proton-proton and antiproton-proton
 scattering amplitudes, {\em together} with cosmic ray data (using Glauber  
theory), to predict proton--air
 and proton-proton cross sections at energies near $\sqrt s \approx$  
30 TeV.
The p-air cosmic ray
 measurements greatly reduce the errors in the high energy
proton-proton and proton-air cross section predictions---in turn,  
greatly reducing the errors in the fit parameters. From this analysis,  
we can then compute the survival probability of rapidity gaps in high  
energy $\pbar p$ and $pp$ collisions, with high accuracy in a quasi  
model-free environment. Using an additive quark model and vector meson  
dominance, we note that that the survival probabilities are identical,  
at the {\em same} energy, for $\gamma p$ and $\gamma \gamma$  
collisions, as well as for nucleon-nucleon collisions.  Significantly,  
our analysis finds large values for  gap survival probabilities,  
$\approx 30$\% at $\sqrt s=200$ GeV, $\approx 21$\% at $\sqrt s=1.8$  
TeV and $\approx 13$\% at $\sqrt s=14$ TeV.
\end{abstract}
\end{titlepage} 
%
\pagenumbering{arabic}
\renewcommand{\thepage}{-- \arabic{page}\ --}  
%

Rapidity gaps are an important tool in new-signature physics for  
ultra-high energy $\pbar p$ collisions. In this note, we will use an  
eikonal model to make a reliable calculation of the survival  
probability of rapidity gaps in nucleon-nucleon collisions.


In an eikonal model\cite{blockhalzenpancheri}, we define our (complex)  
eikonal $\chi(b,s)$ so that $a(b,s)$, the (complex) scattering  
amplitude in impact parameter space $b$, is given by
\be
a(b,s)=\frac{i}{2}\left(1-e^{i\chi(b,s)}\right)
=\frac{i}{2}\left(1-e^{-\chii(b,s)+i\chir(b,s)}
\right).\label{eik}
\ee
Using the optical theorem,
the total cross section $\sigma_{\rm tot}(s)$ is given by
\begin{equation}
\sigma_{\rm tot}(s)=2\int\,\left[1-e^{-\chii  
(b,s)}\cos(\chir(b,s))\right]\,d^2\vec{b},\label{sigtot}
\end{equation}
the elastic scattering cross section
$\sigma_{\rm el}(s)$ is given by
\begin{eqnarray}
\sigma_{\rm  
elastic}(s)&=&\int\left|1-e^{-\chii(b,s)+i\chir(b,s)}\right|^2\,d^2\vec{b}\label{sigel}
\end{eqnarray}
and the inelastic cross section, $\sigma_{\rm inelastic}(s)$, is given by
\be
\sigma_{\rm inelastic}(s)=\sigma_{\rm tot}(s)-\sigma_{\rm  
elastic}(s)=\int\,\left [1-e^{-2\chii(b,s)}\right  
]\,d^2\vec{b}.\label{sigin}
\ee
The ratio of the real to the imaginary part of the forward nuclear  
scattering amplitude, $\rho$,
is given by
\begin{eqnarray}
\rho(s)&=&\frac{{\rm Re}\left\{i(\int  
1-e^{-\chii(b,s)+i\chir(b,s)})\,d^2\vec{b}\right\}}
{{\rm Im}\left\{i(\int  
(1-e^{-\chii(b,s)+i\chir(b,s)})\,d^2\vec{b}\right\}}\label{rho}
\end{eqnarray}
and the nuclear slope parameter $B$ is given by
\be
B=
\frac{\int\,b^2a(b,s)\,d^2\vec{b}}{2\int\,a(b,s)\,d^2\vec{b}}.\label{Bsimple}
\ee

It is readily shown, from unitarity and \eq{sigin}, that the  
differential probability in impact parameter space $b$, for {\em not}  
having an inelastic interaction, is given by
\be
P_{\rm no\  inelastic}=e^{-2\chii(b,s)}.\label{noinelastic}
\ee

Our even QCD-Inspired eikonal $\chi_{\rm even}$ is given by the sum of  
three contributions, gluon-gluon, quark-gluon and quark-quark, which  
are individually factorizable into a product of a cross section   
$\sigma (s)$ times an impact parameter space distribution function  
$W(b\,;\mu)$,  {\em i.e.,}:
\begin{eqnarray}
 \chi^{\rm even}(s,b)& = &\chi_{\rm gg}(s,b)+\chi_{\rm  
qg}(s,b)+\chi_{\rm qq}(s,b)\nonumber\\
&=&i\left[\sigma_{\rm gg}(s)W(b\,;\mu_{\rm gg})+\sigma_{\rm  
qg}(s)W(b\,;\sqrt{\mu_{\rm qq}\mu_{\rm gg}})+\sigma_{\rm  
qq}(s)W(b\,;\mu_{\rm qq})\right],\label{eq:chieven}
\end{eqnarray}
where the impact parameter space distribution function is the  
convolution of a pair of dipole form factors:
\begin{equation}
W(b\,;\mu)=\frac{\mu^2}{96\pi}(\mu b)^3K_3(\mu b).\label{W}
\end{equation}
It is normalized so that $\int W(b\,;\mu)d^2 \vec{b}=1.$ Hence, the  
$\sigma$'s in \eq{eq:chieven} have the dimensions of a cross section.  
The factor $i$ is inserted in \eq{eq:chieven} since the high energy  
eikonal is largely imaginary (the $\rho$ value for nucleon-nucleon  
scattering is rather small).
The total even contribution is not yet analytic.
For large $s$, the {\rm even} amplitude in \eq{eq:chieven} is made  
analytic by the substitution (see  the table on p. 580 of reference   
\cite{bc}, along with reference \cite{eden}) $s\rightarrow  
se^{-i\pi/2}.$ The quark contribution $\chi_{\rm qq}(s,b)$ accounts for  
the constant cross section and a Regge descending component ($\propto  
1/\sqrt s$), whereas the mixed quark-gluon term $\chi_{\rm qg}(s,b)$  
simulates diffraction ($\propto \log s$).  The gluon-gluon term  
$\chi_{\rm gg}(s,b)$, which eventually rises as a power law   
$s^\epsilon$,  accounts for the rising cross section and dominates at  
the highest energies. In \eq{eq:chieven}, the inverse sizes (in impact  
parameter space) $\mu_{\rm qq}$ and $\mu_{\rm gg}$ are to be fit by  
experiment, whereas the quark-gluon inverse size is taken as  
$\sqrt{\mu_{\rm qq}\mu_{\rm gg}}$. For more detail, see  
ref.\cite{blockhalzenpancheri}.

The high energy analytic {\em odd} amplitude (for its structure in  
$s$, see eq. (5.5b) of reference \cite{bc}, with $\alpha =0.5$)
that fits the data is given by
\begin{eqnarray}
\chii^{\rm odd}(b,s)&=&-\sigma_{\rm odd}\,W(b;\mu_{\rm  
odd}),\label{oddanalytic}
\end{eqnarray}
with $\sigma_{\rm odd}\propto 1/\sqrt s$, and
with
\be
W(b,\mu_{\rm odd})=\frac{\mu_{\rm odd}^2}{96\pi}(\mu_{\rm odd}  
b)^3\,K_3(\mu_{\rm odd} b)\label{Woddnormalization},
\ee
normalized so that
$\int W(b\,;\mu_{\rm odd})d^2 \vec{b}=1.$ 
Hence, the $\sigma_{\rm odd}$ in \eq{oddanalytic} has the dimensions  
of a cross section.
Finally,
\be
\chi^{\pbar p}_{pp}=\chi_{\rm even}\pm \chi_{\rm odd}.\label{totalchi}
\ee

\begin{figure}[h,t,b] 
\begin{center}
\mbox{\epsfig{file=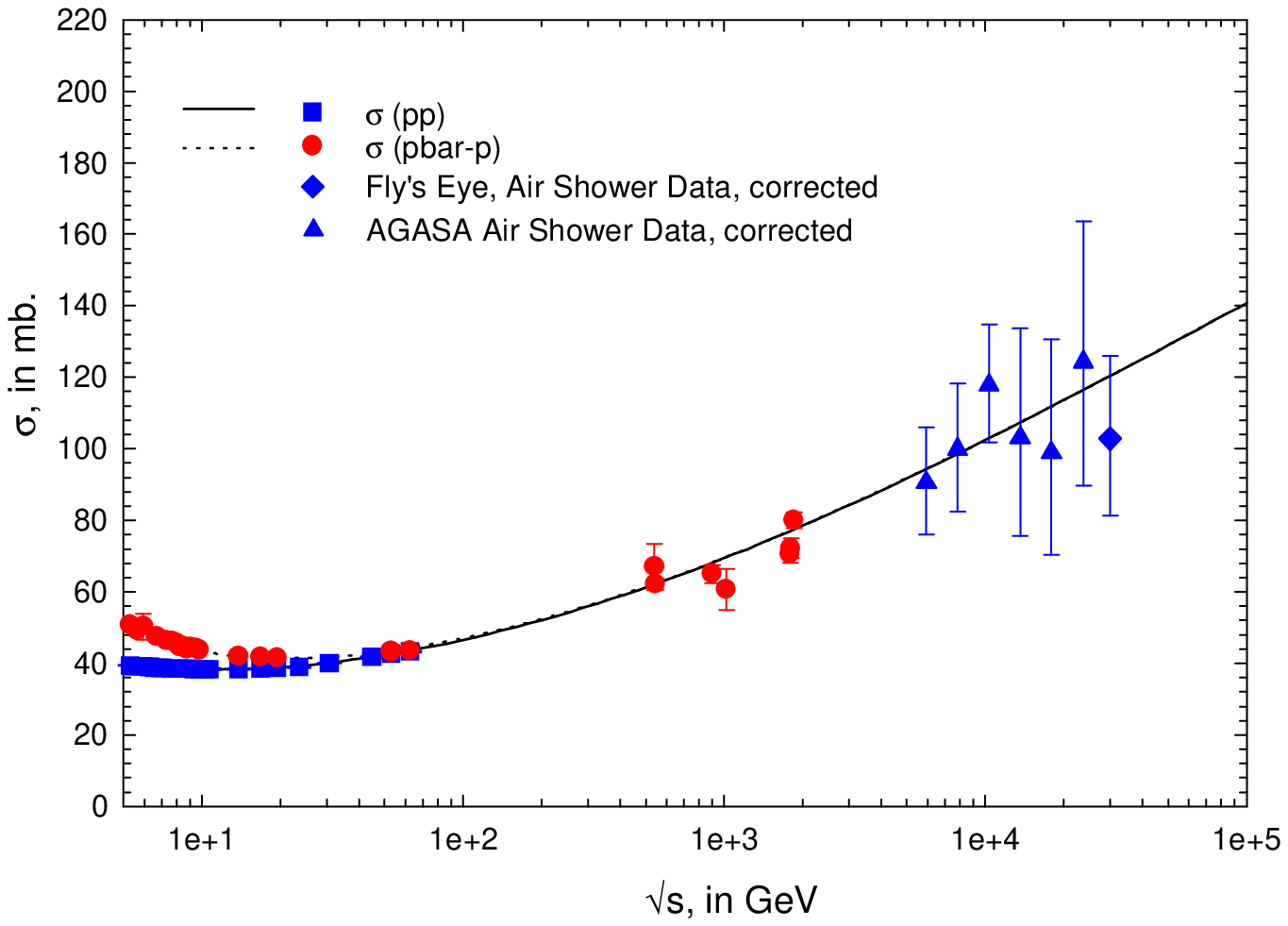%
,
width=6in,bbllx=75pt,bblly=370pt,bburx=525pt,bbury=675pt,clip=%
}}
\end{center}
\caption[]{ \footnotesize
 The fitted $\sigma_{pp}$ and $\sigma_{\bar pp}$, in mb {\em vs.}  
$\sqrt s$, in GeV, for the QCD-inspired fit of  total cross section,  
$B$ and $\rho$  for both $pp$ and $\bar pp$. The accelerator data  
(squares are pp and circles are $\bar {\rm   p}$p ) and  the cosmic ray  
points (diamond, Fly's Eye and triangles, AGASA) have been fitted  
simultaneously. The cosmic ray data that are shown have been converted  
from  $\spai$ to $\sigma_{pp}$.}
\label{fig:ppcurves}
\end{figure}

\begin{figure}[h] 
\begin{center}
\mbox{\epsfig{file=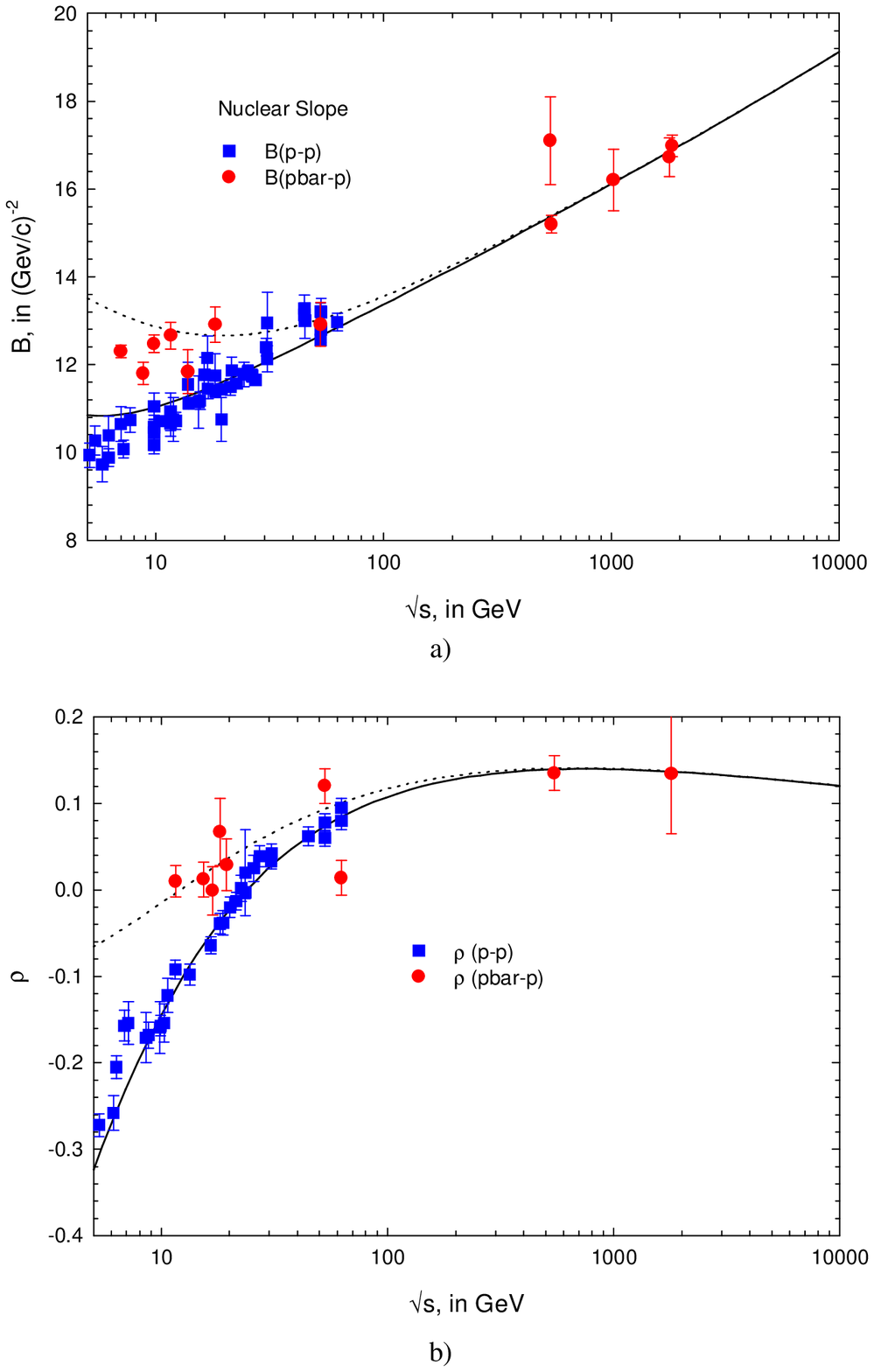%
            ,width=6in,bbllx=56pt,bblly=119pt,bburx=512pt,bbury=730pt,clip=%
}}
\end{center}
\caption[]{ \footnotesize
 The fitted values for the nuclear slope parameters $B_{pp}$ and  
$B_{\bar p p}$,  in (GeV/c)$^{-2}$ {\em vs.} $\sqrt s$, in GeV, for the  
QCD-inspired fit are shown in (a). In (b), the fitted values for  
$\rho_{\bar pp}$ and $\rho_{pp}$ are shown.}
\label{fig:bandrho}
\end{figure}

We have constructed a QCD-inspired parameterization of the forward
 proton--proton and proton--antiproton scattering amplitudes\,\cite{block}
 which is analytic, unitary, satisfies crossing symmetry and, using a  
$\chi^2$ procedure, fits all accelerator data\cite{orear} of
$\sigma_{\rm tot}$, nuclear slope parameter $B$
 and $\rho$, the ratio of the real-to-imaginary part of the forward
 scattering amplitude for both $pp$ and $\pbar p$ collisions; see  
Fig.\,\ref{fig:ppcurves} and Fig.\,\ref{fig:bandrho}. In addition, the  
high energy cosmic ray cross sections of Fly's Eye\,\cite{fly} and
AGASA\,\cite{akeno} experiments are also simultaneously  
used\cite{nonglobal}.  %
Because our parameterization is
 both unitary and analytic, its high energy predictions are
 effectively model--independent, if you require that the proton is
asymptotically a black disk. Using vector meson
 dominance and the additive quark model, we find further support for  
our QCD
fit---it accommodates a wealth of
 data on photon-proton and photon-photon interactions without the
 introduction of new parameters\cite{blockhalzenpancheri}.  A {\em  
major} difference between the present
result, in which we simultaneously  fit the cosmic ray and accelerator  
data,
and our earlier result\cite{nonglobal}, in which only accelerator data are  
used, is a {\em significant} reduction (about a factor of 2.5) in the  
$\approx 1.5$ \% error of our prediction for
$\sigma_{pp}$ at $\sqrt s=30 $ TeV, which results from major  
reductions in our fit parameters.

The plot of $\sigma_{pp}$ {\em vs.} $\sqrt s$, including the  cosmic
ray data that have been converted from $\spai$ to $\sigma_{pp}$, is  
shown in Fig.\,\ref{fig:ppcurves}. %
Clearly,
 we have an excellent fit, with good agreement between AGASA and Fly's  
Eye. The overall agreement between the accelerator and the cosmic
ray $pp$ cross sections with our QCD-inspired fit, as shown in
Fig.\,\ref{fig:ppcurves}, is striking.
%

As an example of a large rapidity gap process, we consider the  
production cross section for Higgs-boson production through W fusion.   
The inclusive differential cross section in impact parameter space $b$  
is given by %
$\frac{d\sigma}{d^2\vec{b}}=\sigma_{{\rm WW}\rightarrow {\rm  
H}}\,W(b\,;\mu_{\rm qq}),$ 
where we have assumed that $W(b\,;\mu_{\rm qq})$ (the differential  
impact parameter space {\em quark} distribution in the proton) is the  
same as that of the W bosons.

The cross section for producing the Higgs boson {\em and} having a  
large rapidity gap (no secondary particles) is given by
\be
\frac{d\sigma_{\rm gap}}{d^2\vec{b}}=\sigma_{{\rm WW}\rightarrow {\rm  
H}}\,W(b\,;\mu_{\rm qq})e^{-2\chii(s,b)}=\sigma_{{\rm WW}\rightarrow  
{\rm H}}\,\frac{d(|S|^2)}{d\vec{b}^2}. \label{eq:nosecondaries}
\ee
In \eq{eq:nosecondaries} we have used \eq{sigin} to get the  
exponential suppression factor, and have used the normalized impact  
parameter space distribution $W(b\,;\mu_{\rm qq})=\frac{\mu_{\rm  
qq}^2}{96\pi}(\mu_{\rm qq} b)^3K_3(\mu_{\rm qq} b)$, with $\mu_{\rm  
qq}=0.901\pm 0.005$ GeV, whose error comes from the fitting  
routine\cite{nonglobal}.

We now generalize and define  $<|S|^2>$, the survival probability of  
{\em any} large rapidity gap, as
\be
<|S|^2>=\int W(b\,;\mu_{\rm  
qq})e^{-2\chii(s,b)}d^2\,\vec{b}.\label{eq:survival}
\ee
We note that the energy dependence of the survival probability  
$<|S|^2>$ is through the energy dependence of $\chii$, the imaginary  
part of the eikonal.

\begin{figure}[h,t] 
\begin{center}
\mbox{\epsfig{file=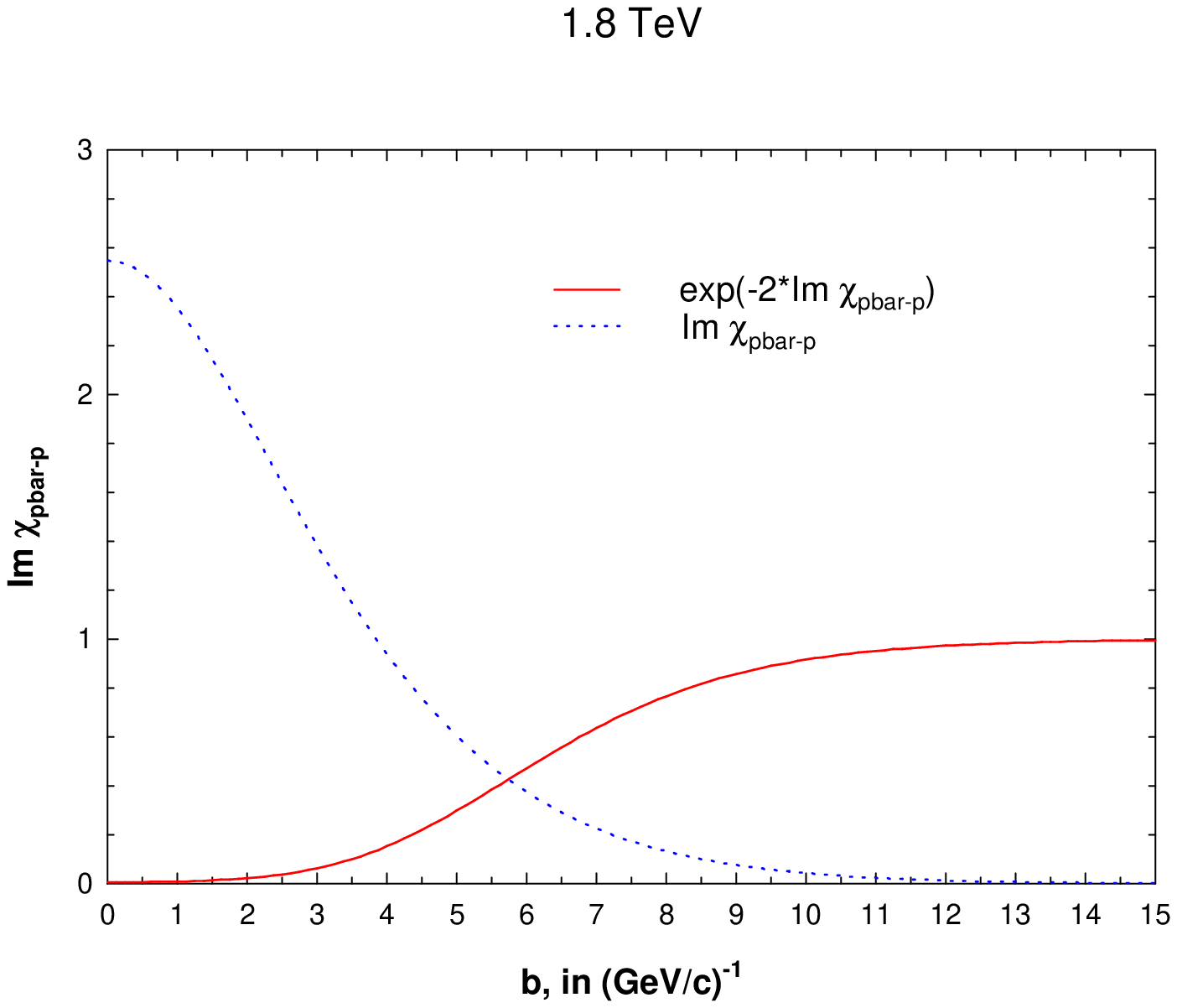%
            ,width=4.3in,bbllx=12pt,bblly=353pt,bburx=437pt,bbury=680pt,clip=%
}}
\end{center}
\caption[]{ \footnotesize
 A plot of the eikonal $Im \chi$ and the exponential damping factor  
$e^{-2*Im \chi}$ for $\pbar p$ collisions, at $\sqrt s=1.8$ TeV {\it  
vs.} the impact parameter $b$, in (GeV/c)$^{-1}$.}
\label{fig:1800}
\end{figure}

For illustration, we show in Fig. \ref{fig:1800}  a plot of ${\rm  
Im}\, \chi_{\bar pp}$ and the exponential damping factor of  
\eq{eq:survival}, as a function of the impact parameter $b$, at  
$\sqrt{s}=1.8$ TeV.
The results of numerical integration of \eq{eq:survival} for the  
survival probability $<S^2>$ at various c.m.s.\ energies   are  
summarized in Table \ref{table:survival}.%
\begin{table}[h]                   
%
\begin{tabular}[b]{|l||c|c|}
      \hline
      C.M.S. Energy (GeV)&Survival Probability($\bar pp$), in \%  
&Survival Probability($pp$), in \%\\ \hline
     63&$37.0\pm 0.9$&$37.5\pm 0.9$\\
     546&$26.7\pm 0.5$&$26.8\pm 0.5$\\
     630&$26.0\pm 0.5$&$26.0\pm 0.5$\\
     1800&$20.8\pm 0.3$&$20.8\pm 0.3$\\
     14000&$\ \  12.6\pm 0.06$&\ \ $12.6\pm 0.06$\\
     40000&$\ \ \ 9.7\pm 0.07$&$\ \ \ 9.7\pm 0.07$\\
     \hline
\end{tabular}
     \vspace{-.2in} \\
     \caption{\footnotesize
The survival probability, $<|S|^2>$, in \%, for $\bar pp$ and $pp$  
collisions, as a function of c.m.s. energy.}\label{table:survival}
\end{table}
As emphasized earlier, the errors in $<|S|^2>$ are quite small, due  
to the fitting parameter errors being very small when we determine the  
eikonal using  both accelerator {\em and} cosmic ray data.

Further, we find for the quark component that the mean squared radius  
of the quarks in the nucleons, $<R_{\rm nn}^2>$, is given by
$<R_{\rm nn}^2>=\int b^2 W(b;\mu_{\rm qq})\,d^2\vec b=16/\mu_{\rm  
qq}^2=19.70\ {\rm GeV}^{-2}$. Thus, $b_{\rm rms}$, the r.m.s. impact  
parameter radius is given by $b_{\rm rms}=4/\mu_{\rm qq}=4.44$  
GeV$^{-1}$. Inspection of Fig. \ref{fig:1800} (1.8 TeV) at $b_{\rm  
rms}$ shows a sizeable probability for no interaction  ($e^{-2\chii}$)  
at that typical impact parameter value.

%

\begin{figure}[h,b] 
\begin{center}
\mbox{\epsfig{file=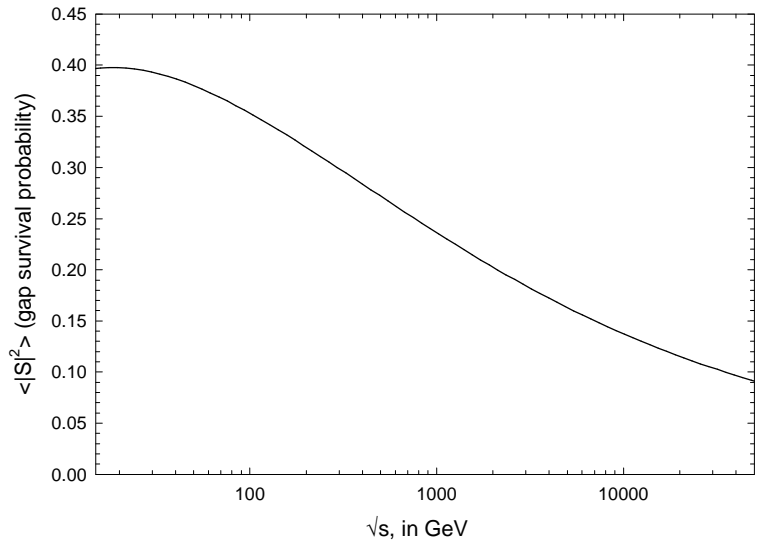%
            ,width=4.5in,bbllx=53pt,bblly=250pt,bburx=525pt,bbury=580pt,clip=%
}}
\end{center}
\caption[]{ \footnotesize
 The energy dependence of $<|S|^2>$, the large rapidity gap survival  
probability  {\em vs.} $\sqrt s$, in GeV.}
\label{fig:survival}
\end{figure}

In ref. \cite{blockhalzenpancheri}, using the additive quark model, we  
show that the eikonal $\chi^{\gamma p}$ for $\gamma p$ reactions is  
found by substituting $\sigma\rightarrow \frac{2}{3}\sigma$,  
$\mu\rightarrow\sqrt{\frac{3}{2}}\mu$ into $\chi^{\rm even}(s,b)$,  
given by \eq{eq:chieven}. In turn, $\chi^{\gamma \gamma}$ for  
$\gamma\gamma$ reactions is found  by substituting $\sigma\rightarrow  
\frac{2}{3}\sigma$, $\mu\rightarrow\sqrt{\frac{3}{2}}\mu$ into  
$\chi^{\gamma p}(s,b)$.
Thus,
\be
\chi^{\gamma p}(s,b)= i\left[\frac{2}{3}\sigma_{\rm  
gg}(s)W(b\,;\sqrt{\frac{3}{2}}\mu_{\rm gg})+\frac{2}{3}\sigma_{\rm  
qg}(s)W(b\,;\sqrt{\frac{3}{2}}\sqrt{\mu_{\rm qq}\mu_{\rm  
gg}}+\frac{2}{3}\sigma_{\rm qq}(s)W(b\,;\sqrt{\frac{3}{2}}\mu_{\rm  
qq})\right], \label{eq:chigp}
\ee
and
\be
\chi^{\gamma \gamma}(s,b)= i\left[\frac{4}{9}\sigma_{\rm  
gg}(s)W(b\,;\frac{3}{2}\mu_{\rm gg})+\frac{4}{9}\sigma_{\rm  
qg}(s)W(b\,;\frac{3}{2}\sqrt{\mu_{\rm qq}\mu_{\rm  
gg}}+\frac{4}{9}\sigma_{\rm qq}(s)W(b\,;\frac{3}{2}\mu_{\rm  
qq})\right]. \label{eq:chigg}
\ee
It can be shown\cite{me}, that by an appropriate change of variables, that
\be
<|S^{\gamma p}|^2>=\int W(b\,;\sqrt{\frac{3}{2}}\mu_{\rm  
qq})e^{-2\chii^{\gamma p}(s,b)}
d^2\,\vec{b}\label{eq:survivalgp}
\ee
where $\chi^{\gamma p}(s,b)$ is given by \eq{eq:chigp},
and
\be
<|S^{\gamma \gamma}|^2>=\int W(b\,;\frac{3}{2}\mu_{\rm  
qq})e^{-2\chii^{\gamma \gamma}(s,b)}
d^2\,\vec{b},\label{eq:survivalgg}
\ee
where $\chi^{\gamma \gamma}(s,b)$ is given by \eq{eq:chigg}, are {\em  
both} equal to
\be
<|S^{\rm nn}|^2>=\int W(b\,;\mu_{\rm qq})e^{-2\chii^{\rm even}(s,b)}
d^2\,\vec{b}\label{eq:survivalnn}
\ee
where $\chi^{\rm even}(s,b)$ is given by \eq{eq:chieven}.  Thus, we  
have the interesting result that
\be
<|S^{\gamma p}|^2>=<|S^{\gamma \gamma}|^2>=<|S^{\rm  
nn}|^2>.\label{eq:allequal}
\ee
Neglecting the small differences at low energy between $\pbar p$ and  
$pp$ collisions, we see from \eq{eq:allequal} that $<|S|^2>$, the  
survival probability for either nucleon-nucleon, $\gamma p$ and $\gamma  
\gamma$ collisions, is {\em reaction-independent}, depending {\em  
only} on $\sqrt{s}$, the cms energy of the collision.  We show in ref.  
\cite{me} that this result is true for {\em any} factorization scheme  
where the eikonal factorizes into sums of $\sigma_i (s)\times  
W_i(b;\mu)$, with the scaling feature that the product  
$\sigma_i\mu_i^2$ is reaction-independent---not only for the additive  
quark model that we have employed here.  The energy dependence of the  
large rapidity gap survival probability $<|S|^2>$ calculated from  
\eq{eq:survivalnn} is given in
Fig. \ref{fig:survival}.

This somewhat surprising result can be more readily understood once one realizes  
that the survival probability is a function of the product $\sigma \mu^2$ and the dimensionless variable 
$x = \mu_{\rm qq} b$. This is most simply seen in a toy model in which the even eikonal is given by
\be 
\chi^{\rm even}(s,b)_{\rm toy}=\sigma_{\rm qq}W(b;\mu_{\rm qq})=
\sigma_{\rm qq}\mu_{\rm qq}^2\times\frac{(\mu_{\rm qq}b)^3K_3(\mu_{\rm qq}b)}{96\pi}=
\sigma_{\rm qq}\mu_{\rm qq}^2\times \frac{x^3K_3(x)}{96\pi},\label{eq:toychi} 
\ee
and hence,
\be
<|S^{\rm nn}|_{\rm toy}^2>= \frac{1}{96\pi}
\int e^{-2\sigma_{\rm qq}\mu_{\rm qq}^2 x^3K_3(x)/96\pi}d^2\vec x.\label{eq:toysurvival}
\ee
In our scheme  
$\sigma_{\rm qq}\mu_{\rm qq}^2$ is the same for nucleon-nucleon, $\gamma p$ and $\gamma \gamma$ interactions, since cross sections are multiplied by $\sqrt{\frac{2}{3}}$ and $\frac{2}{3}$ for $\gamma p$ and $\gamma\gamma$, respectively, whereas $\mu_{\rm qq}^2$ is divided by $\sqrt{\frac{2}{3}}$ and $\frac{2}{3}$. Although  
$x$ is different for the 3 processes---being $ \mu_q b$ for nucleon-nucleon,  $\sqrt{\frac{3}{2}}\mu_{\rm qq}b$ for $\gamma p$ and $\frac{3}{2}\mu_{\rm qq}b$ for $\gamma \gamma$---it only plays the role of an  
integration variable and therefore the dependence on the subprocess  
label disappears. We see from \eq{eq:toysurvival} that  S=S($\sigma_{\rm qq}\mu_{\rm qq}^2$), which is {\em process-independent}.  This argument is easily generalized to the full eikonals of  \eq{eq:chieven}, \eq{eq:chigp} and \eq{eq:chigg}.

The physics is now  
clear. One could be tempted to conclude that the survival probability  
is larger for $\gamma p$ than for $pp$ interactions because there are  
only 2 quarks in the photon and 3 in the proton to produce additional  
inelastic collisions filling the gap. And that is true and it is  
reflected by the factor 2/3 change in the cross section. This is not  
the whole story, however. In the eikonal model, the inverse transverse size of  
the 2-quark system (photon) is larger than that of the proton (the $\frac{3}{2}\mu_{\rm qq}^2$ factor) and the two effects compensate. Therefore, we have the same $<|S|^2>$ for all processes.


We have been able to calculate the survival probability $<|S|^2>$ to a  
high degree of accuracy by using an eikonal that has been found by  
fitting both accelerator and cosmic ray data over a very large energy  
range which includes the LHC energy.  Our numerical results are  
considerably larger than other   
calculations\cite{Maor,Fletcher,Bjorken}.  In the case of ref.  
\cite{Maor} and ref. \cite{Bjorken} it is probably due to their using a  
Gaussian probability distribution in impact parameter space, whereas  
our distribution,
$W(b,\mu_{qq})=\frac{\mu_{qq}^2}{96\pi}(\mu_{qq} b)^3\,K_3(\mu_{qq}  
b)$ which is the Fourier transform of the square of a dipole  
distribution, has a long exponential tail $e^{-\mu b}$, significantly  
increasing the probability of survival.  In the case of ref.  
\cite{Fletcher}, the authors determine the parameters for their minijet  
model using only the Tevatron results. Our large values are more in  
line with the earlier predictions of Gotsman {\em et al}\cite{Gotsman}  
for what they called Regge and Lipatov1 and Lipatov2 models, although  
with somewhat different energy dependences than that shown in Table 1.  
The color evaporation model of \`Eboli {\em et al}\cite{Eboli} gives  
somewhat larger values than our calculation, but again with a different  
energy dependence. Most recently, Khoze {\em et al}\cite{Martin},  
using a two-channel eikonal, have calculated the survival probabilities  
for rapidity gaps in single, central and double diffractive processes  
at several energies, as a function of $b$, the slope of the  
Pomeron-proton vertex. For double diffraction, they have a large range  
of possible parameters. Choosing  $2b=5.5$ GeV$^{-2}$ (corresponding to  
the slope of the electromagnetic proton form factor),  they obtain  
$<|S|^2>=0.26$, 0.21 and 0.15 at $\sqrt {s}=0.54$, 1.8 and 14 TeV,  
respectively. These survival probabilities are in excellent agreement  
with our values given in Table 1. However, their  calculations for  
other choices of $2b$ and for single and central diffractive processes  
do not agree with ours, being model-dependent, with their results  
varying considerably with their choice of parameters and model.

We see that there is a serious model dependence, both in the size of  
the survival probabilities and in their energy dependence. Further,  
until now, there has been no estimates for gap survival probabilities  
for $\gamma p$ and $\gamma\gamma$ reactions. Thus, we hope that our  
quasi model-independent fit to experimental data on $\bar p p$ and $pp$  
total cross sections, $\rho$ values and nuclear slopes $B$, over a  
large energy range, $\sqrt s$ = 15 GeV  to 30,000 GeV, provides a  
reliable quantitative estimate of the survival probability $<|S|^2>$ as  
a function of energy, for both $\pbar p$, $pp$, $\gamma p$ and $\gamma  
\gamma$ collisions.  The fact that our estimates of large rapidity gap  
survival probabilities are independent of reaction, thus being equal  
for nucleon-nucleon, $\gamma p$ and $\gamma \gamma$ processes---the  
equality surviving any particular factorization scheme---has many  
interesting experimental consequences.
%
%

%
%
\end{document}